\documentclass[a4paper,12pt,onecolumn,oneside,notitlepage,final]{article}
\usepackage[dvipdfmx]{graphicx}
\usepackage{amssymb} %
\usepackage{authblk}
\def\Journal#1#2#3#4{{#1} {\bf #2}, #3 (#4)}

\def\APJS{Astrophys. J. Suppl.}

\def\JCAP{J. Cosmol. Astropart. Phys.}

\def\JCAP{JCAP}

\def\NPA{Nucl. Phys. A}
\def\NPB{Nucl. Phys. B}

\def\PLA{Phys. Lett. A}
\def\PLB{Phys. Lett. B}

\def\PLBOLD{Phys. Lett.}

\def\PRD{Phys. Rev. D}

\def\RPP{Rep. Prog. Phys.}

\begin{document}
\title{Fermi-Boltzmann statistics of neutrinos and relativistic effective degrees of freedom in the early universe}

\author{Jun Iizuka and Teruyuki Kitabayashi\footnote{teruyuki@keyaki.cc.u-tokai.ac.jp}}

\affil{Department of Physics, Tokai University,\\
4-1-1 Kitakaname, Hiratsuka, Kanagawa, 259-1292, Japan\\
}

\date{\small \today}
\maketitle

\begin{abstract}
We investigate the effect of the presence of non-pure fermionic neutrinos on the relativistic effective degrees of freedom in the early universe. The statistics of neutrinos is transformed continuously from Fermi-Dirac to Maxwell-Boltzmann statistics. We find that the relativistic degrees of freedom decreases with the deviation from pure Fermi-Dirac statistics of neutrinos if there are constant and large lepton asymmetries. Additionally, we confirm that the change of the statistics of neutrinos from Fermi-Dirac to Maxwell-Boltzmann is not sufficient to cover the excess of the effective number of neutrinos.
\end{abstract}

\section{Introduction}
\label{sec:intro}
Neutrinos have spin $1/2$ and they are classified as fermions. The electrons have spin $1/2$ and possible violation of Fermi-Dirac statistics for the electrons is strongly restricted by experiments \cite{Bartalucci2006PLB}. It is natural that we understand the neutrinos obey purely Fermi-Dirac statistics on the analogy of the electrons. However, the possibility of the non-pure fermionic neutrinos is not excluded perfectly. Although the possibility of the pure bosonic neutrinos (they obey purely Bose-Einstein statistics \cite{Dolgov2005PLB}) is excluded by double-beta decay \cite{Barabash2007PLB}, it does not mean that the neutrinos are pure fermions. Moreover, properties of neutrinos are so different from other spin one half particles, e.g., tiny masses, large mixings, electrically neutral. Thus, the Pauli exclusion principle may be violated for neutrinos and they may possess mixed statistics \cite{Choubey2006PLB,Ignatiev2006PLA,Tornow2010NPA,Vergados2012RPP,NEMO2014NPA}.

The thermal history of the early universe is affected by the change of the statistics of neutrinos. Dolgov, et.al. introduced the Fermi-Bose parameter to study the effects of continuous transition from Fermi-Dirac to Bose-Einstein statistics of neutrinos and discussed the possible modification of the big bang nucleosynthesis (BBN) in the presence of non-pure fermionic neutrinos \cite{Dolgov2005JCAP}.

In the main part of this paper, we investigate the effect of the presence of non-pure fermionic neutrinos on the relativistic effective degrees of freedom at temperature $T=0.5 - 500$ GeV (radiation dominant era, before BBN). The assumption of violation of the Pauli principle makes problems but we put aside discussion of these problems \cite{Dolgov2005PLB} and we restrict ourselves to the phenomenological approach. The large lepton asymmetries are assumed at GeV scale in our scenario. The large lepton asymmetries at MeV scale are almost excluded by the standard BBN cosmology \cite{Mangano2012PLB,Ichimasa2014PRD}. However, the large lepton asymmetries at GeV scale are compatible with current observations \cite{Stuke2012JCAP}. The relation between the relativistic degrees of freedom and the large lepton asymmetries at GeV scale has already discussed by Stuke, et.al.\cite{Stuke2012JCAP}, although, neutrinos are pure fermions in their analysis. The research by Dolgov, et.al.,\cite{Dolgov2005JCAP} as well as by Stuke, et. al.,\cite{Stuke2012JCAP} impressed us. In this paper, we perform the similar analysis as Ref.\cite{Stuke2012JCAP} but we introduce the Fermi-Bose parameter \cite{Dolgov2005JCAP} to relax the statistics of neutrinos. 

The energy density of the ultra-relativistic bosons is larger than that of the ultra-relativistic fermions \cite{Kolb1990}. One may expect that the energy density of the universe in the radiation dominant era increases with deviation of the statistics of neutrinos from pure Fermi-Dirac case. In spite of this naive expectation, we show that the relativistic degrees of freedom decreases with deviation from pure Fermi-Dirac statistics of neutrinos if there are constant and large lepton asymmetries at GeV scale universe.

We also study the bounds on the chemical potentials of neutrino $\mu_\nu$ or degenerate parameter of neutrinos $\xi_\nu=\mu_\nu/T$ as a function of the Fermi-Bose parameter at temperature $T\simeq 1$ MeV (around BBN). These bounds on the degenerate parameter are derived from BBN, more precisely the effective number of neutrinos $N_{eff}$ at BBN. We show that quite large degenerate parameter is required if we try to explain the excess of the effective number of neutrinos, $N_{eff}>3$, without non-standard model particles. 

\section{Statistics of neutrinos and relativistic effective degrees of freedom}
The distribution function of particle species $i$ is given by
\begin{eqnarray}
f_i=\frac{g_i}{e^{(E-\mu_i)/T}+\kappa_i},
\label{Eq:f}
\end{eqnarray}
where $g_i$, $E$, $\mu_i$ and $T$ denote number of internal degrees of freedom, energy, chemical potential and temperature of particle $i$, respectively \cite{Dolgov2005JCAP}. In this paper, we can regard the temperature $T$ as the temperature of the thermal bath (temperature of the photons). The Fermi-Bose parameter $\kappa_i$ describes the statistics of particle $i$; $\kappa_i=1$ and $\kappa_i=-1$ correspond the pure Fermi-Dirac distribution and the pure Bose-Einstein distribution while $\kappa_i=0$ corresponds the pure Maxwell-Boltzmann distribution. We assume that the expression of distribution function in Eq.(\ref{Eq:f}) remains for $-1 \le \kappa_i \le 1$ \cite{Dolgov2005JCAP}. We also assume that the spin one half particles $i$ obey "Fermi-Boltzmann" distribution with $0\le \kappa_i \le 1$ while spin integer particles $i$ obey "Bose-Boltzmann" distribution with $-1\le \kappa_i \le 0$. With these assumptions, the Fermi-Bose parameter $\kappa_i$ describes the continuous transition either from Fermi-Dirac to Maxwell-Boltzmann or from Bose-Einstein to Maxwell-Boltzmann distributions.

The number density $n_i$, energy density $\rho_i$, pressure $P_i$ and entropy density $s_i$ of particle species $i$ are obtained as follows \cite{Kolb1990}:
\begin{eqnarray}
n_i&=&\frac{1}{2\pi^2}\int_{m_i}^\infty E(E^2-m_i^2)^{1/2}f_i dE,  \label{Eq:n} \\
\rho_i&=&\frac{1}{2\pi^2}\int_{m_i}^\infty E^2(E^2-m_i^2)^{1/2}f_i dE, \label{Eq:rho} \\
P_i&=&\frac{1}{6\pi^2}\int_{m_i}^\infty (E^2-m_i^2)^{3/2}f_i dE, \label{Eq:P} \\
s_i&=&\frac{\rho_i+P_i-\mu_i n_i}{T}, \label{Eq:s}
\end{eqnarray}
where $m_i$ denotes the mass of particle $i$. There are other expressions of the entropy density 
\begin{eqnarray}
s_i=-\frac{1}{(2\pi)^3} \int_{-\infty}^{\infty} [f_i\ln f_i + \kappa_i (1-\kappa_i f_i)\ln(1-\kappa_i f_i)] d^3p,
\label{Eq:s_f_FB}
\end{eqnarray}
for pure Fermi-Dirac ($\kappa_i=1$) or pure Bose-Einstein ($\kappa_i=-1$) and 
\begin{eqnarray}
s_i=-\frac{1}{(2\pi)^3}\int_{-\infty}^{\infty} (f_i\ln f_i -f_i) d^3p,
\label{Eq:s_f_MB}
\end{eqnarray}
for pure Maxwell-Boltzmann statistics where $p$ denotes momentum \cite{Bernstein1988}. If the deviation from pure Fermi-Dirac or pure Bose-Einstein statistics is small ($\vert \kappa_i \vert \sim 1$), Eq.(\ref{Eq:s_f_FB}) will be good choice to parametrize the entropy density for Fermi-Boltzmann or Bose-Boltzmann statistics.  However, we would like to assign $\kappa_i=0$ for the Maxwell-Boltzmann case, but Eq.(\ref{Eq:s_f_FB}) with $\kappa_i=0$ is not same as Eq.(\ref{Eq:s_f_MB}). Thus, we use Eq.(\ref{Eq:s}) to estimate the entropy density. 

For the massless and non-degenerate ($\mu_i=0$) pure fermionic ($\kappa_i=1$) and pure bosonic ($\kappa_i=-1$) particles, the total energy density is given by
\begin{eqnarray}
\rho^{m_i=\mu_i=0}=\sum_i\rho_i^{m_i=\mu_i=0} = \frac{\pi^2 T^4}{30} g_*^{m_i=\mu_i=0}, 
\label{Eq:rho_rela}
\end{eqnarray}
where $g_*^{m_i=\mu_i=0}$ denotes well known relativistic effective degrees of freedom \cite{Kolb1990}
\begin{eqnarray}
g_*^{m_i=\mu_i=0}=\sum_{i=bosons}g_i+\frac{7}{8}\sum_{i=fermions}g_i.
\label{Eq:gast_rela}
\end{eqnarray}
According to Stuke, et.al.\cite{Stuke2012JCAP}, we define the general relativistic effective degrees of freedom $g_*=g_*(m_i, \kappa_i,\mu_i)$ for any $m_i$, $\kappa_i$ and $\mu_i$ as follows
\begin{eqnarray}
g_*=\frac{30}{\pi^2T^4}\rho,
\label{Eq:gast_general1}
\end{eqnarray}
where $\rho=\sum_i \rho_i(m_i, \kappa_i,\mu_i)$.

To estimate the total energy density $\rho$, the chemical potentials of all particle species $\mu_i$ should be determined. We use the strategy for determining these chemical potentials by Stuke, et.al.\cite{Stuke2012JCAP}. The magnitude of the chemical potential of particle $i$ and its antiparticle $\bar{i}$ are same but the sign of these are opposite, $\mu_i=-\mu_{\bar{i}}$. The beta-decay of down quark $d$ via weak interactions $d \rightarrow u + f + \bar{\nu}_f$ provides $\mu_u+\mu_f = \mu_d+\mu_{\nu_f}$ where $f=e,\mu,\tau$. We assume that $\mu_u=\mu_c=\mu_t$ and $\mu_d=\mu_s=\mu_b$ for quarks. All gauge bosons and Higgs boson have vanishing chemical potentials $\mu_\gamma=\mu_W=\mu_Z=\mu_g=\mu_H=0$. There are only five independent chemical potentials and we take these as $\mu_{\nu_e}, \mu_{\nu_\mu}, \mu_{\nu_\tau}, \mu_u, \mu_d$. These five independent chemical potentials are uniquely determined by the following five conservation laws: 
\begin{eqnarray}
sq&=&-\sum_{i=e,\mu,\tau}n_i + \frac{2}{3}\sum_{i=u,c,t}n_i-\frac{1}{3}\sum_{i=d,s,b}n_i, \label{Eq:sq} \\
sb&=&\frac{1}{3}\sum_{i=quarks}n_i, \label{Eq:sb} \\
s\ell_f &=& n_f + n_{\nu_f}, \quad f=e,\mu,\tau, \label{Eq:sell} 
\end{eqnarray}
where $s=\sum_i s_i$ and $q$ denote total entropy density and electric charge of the universe, respectively. Here, baryon number $b$ and lepton flavour number $\ell_f$ are defined as
\begin{eqnarray}
b=\frac{n_b-n_{\bar{b}}}{s}, \quad \ell_f=\frac{n_f-n_{\bar{f}} +n_{\nu_f}-n_{\bar{\nu}_f}}{s},
\end{eqnarray}
where $n_b$ and $n_{\bar{b}}$ denote the number densities of baryons and anti-baryons, respectively. Eqs.(\ref{Eq:sq}), (\ref{Eq:sb}) and (\ref{Eq:sell}) show the electric charge conservation law, baryon number conservation law and lepton flavour conservation law, respectively. As we addressed in the introduction, the large lepton asymmetries at GeV scale are compatible with current observations, so that the large lepton asymmetries are assumed at GeV scale in our scenario. The neutrino oscillations may ensure the equilibration of the lepton flavour numbers ($\ell_e=\ell_\mu=\ell_\tau$) in the early universe \cite{Mangano2012PLB,Dolgov2002NPB,Abazajian2002PRD,Wong2002PRD}. Although, there are interesting research related to the lepton flavour asymmetries ($\ell_e \neq \ell_\mu \neq \ell_\tau$) in the early universe, for example \cite{Stuke2012JCAP}, we assume that all lepton flavour numbers are the same $\ell=\ell_e=\ell_\mu=\ell_\tau$ for sake of simplicity. 

In this paper, we consider the effect of the change of neutrino statistics on the relativistic effective degrees of freedom. The Fermi-Bose parameters $\kappa_i$ of standard model particles $i$ are assigned as follows: $0 \le \kappa_\nu \le 1$ for neutrinos $\nu_f$ (we assume that $\kappa_\nu=\kappa_{\nu_e}=\kappa_{\nu_\mu}=\kappa_{\nu_\tau}$), $\kappa_\ell=\kappa_q=+1$ for charged leptons $\ell$ and quarks $q$,  $\kappa_{\gamma,W,Z,g}=\kappa_H=-1$ for gauge bosons $\gamma,W,Z,g$ and Higgs boson $H$. In this case 
\begin{eqnarray}
g_*=g_*^{m_i=\mu_i=0}+\Delta g_*(\kappa_\nu= 1,\mu_i \neq 0)+\Delta g_*(\kappa_\nu \neq 1, \mu_i \neq 0),
\label{Eq:gast_general2}
\end{eqnarray}
where $g_*^{m_i=\mu_i=0}$ denotes usual relativistic effective degrees of freedom in Eq.(\ref{Eq:gast_rela}) and $\Delta g_*(\kappa_\nu= 1,\mu_i \neq 0)$ denotes the deviation from $g_*^{m_i=\mu_i=0}$ due to the non-zero lepton asymmetries  which has been reported \cite{Stuke2012JCAP}. The 3rd term in R.H.S in Eq.(\ref{Eq:gast_general2}), $\Delta g_*(\kappa_\nu \neq 1, \mu_i \neq 0)$ is the new contribution on the relativistic effective degrees of freedom $g_*$.

\section{Numerical estimation}
\subsection{GeV scale}
To calculate the energy density $\rho_i$ numerically, we use the following Gauss-Laguerre integration method:
\begin{eqnarray}
\int_0^\infty x^\alpha e^{-x} F(x) dx = \sum_{j=1}^N w_j F(x_j),
\label{Eq:GaussLaguerre}
\end{eqnarray}
where $F(x)$ and $w_j$ denote a function of $x$ and "Gauss-Laguerre weights", respectively \cite{Press1992,Schwarz2009JCAP}. In our calculation, $\alpha=0$ and $N=50$ are chosen. In order to use the formula in Eq.(\ref{Eq:GaussLaguerre}) to calculate $\rho_i$, we estimate  the following integral
\begin{eqnarray}
\rho_i=\frac{1}{2\pi^2}\int_{0}^\infty T(Tx+m_i)^2[(Tx+m_i)^2-m_i^2]^{1/2} f_i dx,
\end{eqnarray}
instead of Eq.(\ref{Eq:rho}), where $x=(E-m_i)/T$ and the distribution function to be
\begin{eqnarray}
f_i=\frac{g_i}{e^{(Tx+m_i-\mu_i)/T}+\kappa_i}.
\end{eqnarray}
Similarly, number density $n_i$ and pressure $P_i$ are calculated by
\begin{eqnarray}
n_i&=&\frac{1}{2\pi^2}\int_{0}^\infty T(Tx+m_i)[(Tx+m_i)^2-m_i^2]^{1/2} f_i dx, \\
P_i&=&\frac{1}{6\pi^2}\int_{0}^\infty T[(Tx+m_i)^2-m_i^2]^{3/2} f_i dx.
\end{eqnarray}
The masses of all particles except neutrinos are taken as the particle data group \cite{PDG2012PRD}.  The direct measurements of the masses of neutrinos are not available. The cosmological constraint on neutrino masses are estimated $\sum m_{\nu} < 0.21 - 1.11$ eV  \cite{Giusarma2014PRD}. We assume that all neutrino masses are same $m_\nu=m_{\mu_e}=m_{\nu_\mu}=m_{\nu_\tau}=1$ eV.  We note that the unknown masses of the neutrinos are irrelevant in our study. To obtain the chemical potentials of particles, nonlinear simultaneous equations (\ref{Eq:sq}), (\ref{Eq:sb}) and (\ref{Eq:sell}) are solved by the Broyden's method \cite{Press1992,Schwarz2009JCAP}. From charge neutrality of the universe, we assume $q=0$ in Eq.(\ref{Eq:sq}). The baryon number is fixed to $b=9 \times 10^{-11}$ \cite{Stuke2012JCAP,Bennett2013APJS,Planck2013arXiv} in Eq.(\ref{Eq:sb}). Only the order of magnitude of the baryon number is important in this paper. Since the measurements of the lepton flavour numbers in the early universe are not established, we keep $\ell=\ell_e=\ell_\mu=\ell_\tau$ as free parameters  in Eq.(\ref{Eq:sell}).

\begin{figure}[t]
\begin{center}
\includegraphics[width=9.5cm]{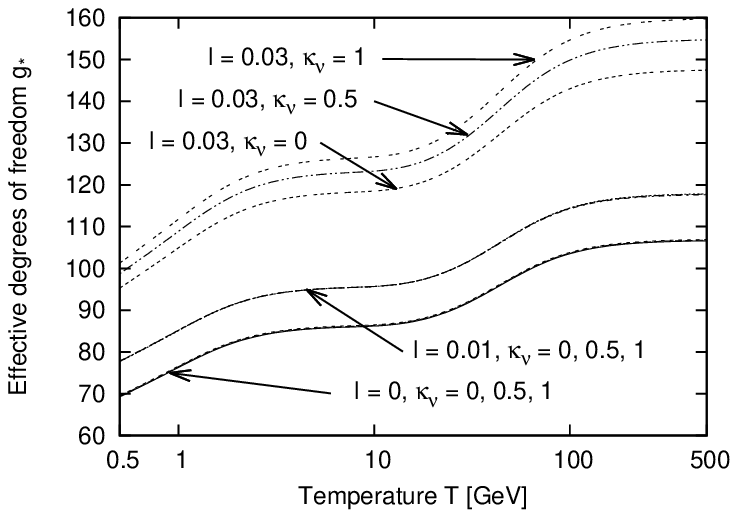}
\includegraphics[width=9.5cm]{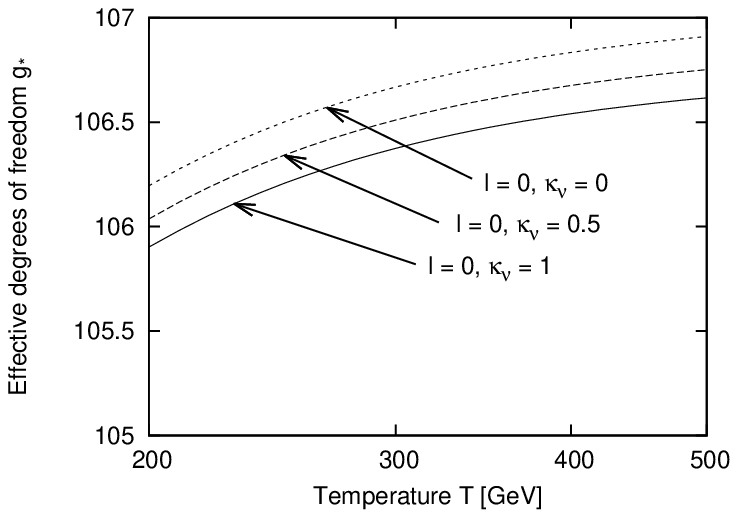}
\caption{Dependence of the relativistic effective degrees of freedom $g_*$ on temperature $T$. The upper figure shows the numerical solution for $\ell=0$, $\ell = 0.01$ and $\ell=0.03$ cases. The lower figure shows the $\ell=0$ case in more detail.}
\label{Fig:gast_T}
\end{center}
\end{figure}

Figure \ref{Fig:gast_T} shows the dependence of the relativistic effective degrees of freedom $g_*$ on temperature $T$. The upper figure shows the numerical solution for $\ell=0$, $\ell = 0.01$ and $\ell=0.03$ cases. There are three curves correspond to $\kappa_\nu=1, 0.5, 0$ for $\ell=0.03$. For $\ell=0.01$ and $\ell=0$, these three curves almost overlap with each other. The lower figure shows the $\ell=0$ case in more detail. Eqs.(\ref{Eq:rho_rela}) and (\ref{Eq:gast_rela}) show that the energy density of pure bosons ($\kappa_i = -1$) is larger than the energy density of pure fermions ($\kappa_i = 1$) by the factor $8/7$ \cite{Dolgov2005JCAP}. Thus one may expect, if neutrinos have non-pure Fermi-Dirac statistics ($\kappa_\nu \neq 1$), their lager energy density would correspond to an increase of the effective degrees of freedom $g_*$ in Eq.(\ref{Eq:gast_general1}). Indeed, if there is no lepton asymmetries (lower figure), relativistic effective degrees of freedom for pure Maxwell-Boltzmann neutrinos case  ($\ell = 0, \kappa _\nu= 0$) is lager than it for pure Fermi-Dirac neutrinos case ($\ell = 0, \kappa_\nu = 1$); $g_*^{\ell = 0, \kappa_\nu=0} > g_*^{\ell = 0,\kappa_\nu=1}$, as our naive expectation. However, if there is large lepton asymmetries, such as $\ell=0.03$  (upper figure), the relativistic effective degrees of freedom for pure Maxwell-Boltzmann neutrinos case ($\ell = 0.03, \kappa_\nu = 0$) is smaller than it for pure Fermi-Dirac neutrinos case ($\ell = 0.03, \kappa_\nu = 1$); $g_*^{\ell=0.03, \kappa_\nu=0} < g_*^{\ell=0.03, \kappa_\nu=1}$.

\begin{figure}[t]
\begin{center}
\includegraphics[width=9.5cm]{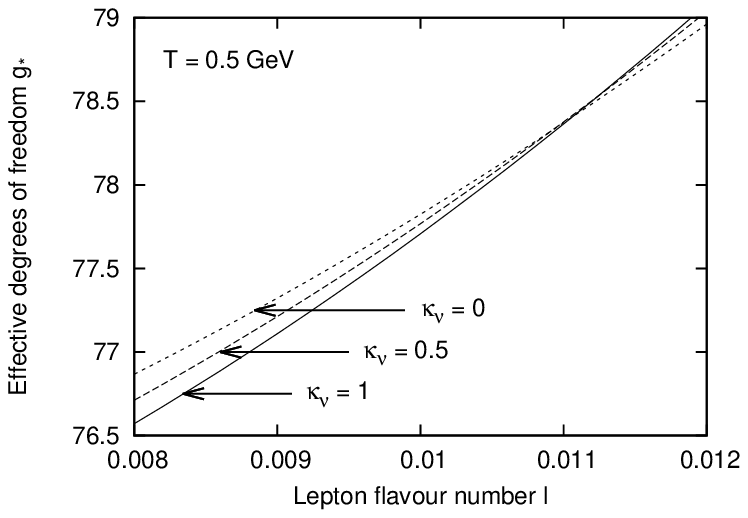}
\includegraphics[width=9.5cm]{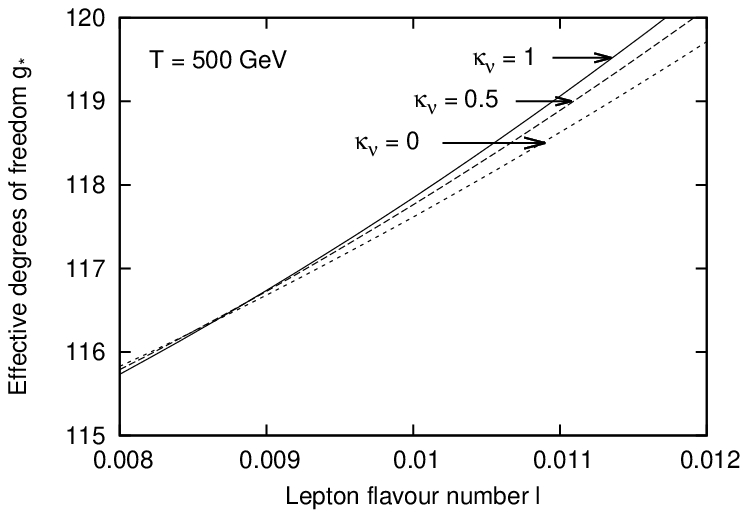}
\caption{The dependence of the relativistic effective degrees of freedom $g_*$ on lepton flavour number $\ell$ for $T=0.5$ GeV (upper) and for $T=500$ GeV (lower). }
\label{Fig:gast_ell}
\end{center}
\end{figure}

Figure \ref{Fig:gast_ell} shows the dependence of the relativistic effective degrees of freedom $g_*$ on lepton flavour number $\ell$. The upper figure shows the result for $T=0.5$ GeV while the lower figure shows the same but for $T=500$ GeV. In the case of $T=0.5$ GeV,  $g_*^{\kappa_\nu=1}$ and $g_*^{\kappa_\nu=0}$ are equivalent at $\ell^{eq} \sim 0.0112$  ($g_*^{\kappa_\nu=1} \lesssim g_*^{\kappa_\nu=0}$ for $\ell \lesssim \ell^{eq}$ and $g_*^{\kappa_\nu=1} \gtrsim g_*^{\kappa_\nu=0}$ for $\ell \gtrsim \ell^{eq}$.) In the case of $T=500$ GeV, the equivalent lepton flavour number $\ell^{eq} \sim 0.0087$ is smaller than it for $T=5$ GeV.  In the range of $0.5 \le$ T[GeV] $\le 500$, the equivalent lepton flavour number to be $0.0087 \lesssim \ell^{eq} \lesssim 0.0112$ and $\ell^{eq}$ decreases with the increasing temperer $T$.

Recall that the relativistic effective degrees of freedom $g_*$ increases with the energy density $\rho$ and this energy density increases with distribution function $f_i=\frac{g_i}{e^{(E-\mu_i)/T}+\kappa_i}$. If there is no lepton asymmetry ($\ell=0$) then the chemical potential of neutrinos vanish $\mu_\nu=0$ and the small $\kappa_\nu$ induces large $f_\nu$ as well as large $g_*$ as our naive expectation; $g_*^{\ell = 0, \kappa_\nu=0} > g_*^{\ell = 0,\kappa_\nu=1}$. Moreover, if there are non-zero lepton asymmetry $\ell \neq 0$, the small $\kappa_\nu$ still induces large $g_*$ as long as the chemical potential $\mu_\nu$ is constant. In other word, if the chemical potential  $\mu_\nu$ is independent of $\kappa_\nu$ and is constant,  the relation of  $g_*^{\mu_\nu = const, \kappa_\nu=0} > g_*^{\mu_\nu = const,\kappa_\nu=1}$ is obtained. We usually consider only pure Fermi-Dirac neutrinos ($\kappa_\nu=1$), so that we usually encounter the constant chemical potential $\mu_\nu$ with constant lepton number $\ell$. 

However, if the Fermi-Bose parameter $\kappa_\nu$ change continuously from $\kappa_\nu=1$ to $\kappa_\nu \neq 1$, even if the lepton flavour number is constant, the chemical potential $\mu_\nu$ is no longer constant with $\kappa_\nu$. The small $\kappa_\nu$ induces large $f_\nu$ while the small $\mu_\nu$ induces small $f_\nu$. In this case the relation of $g_*^{\mu_\nu = const, \kappa_\nu=0} > g_*^{\mu_\nu = const,\kappa_\nu=1}$ is no longer guaranteed and $g_*^{\mu_\nu \neq const, \kappa_\nu=0} < g_*^{\mu_\nu \neq const,\kappa_\nu=1}$ may be allowed. If the effect of $\mu_\nu$ on $f_\nu$ is dominant, the energy density will be reduced. This is the reason why we encounter $g_*^{\ell=0.03, \kappa_\nu=0} < g_*^{\ell=0.03, \kappa_\nu=1}$ in the Figure \ref{Fig:gast_T}.

\begin{table}[t]
\begin{center}
\caption{Chemical potential $\mu_{\nu_e}$ [GeV] and energy density $\rho_{\nu_e}+\rho_{\bar{\nu}_e}$ [GeV] of neutrinos at $T=0.5$ GeV.}
{\begin{tabular}{|c|cc|cc|cc|}
\hline 
         & $\ell=0$   &             & $\ell=0.01$ &                                  & $\ell=0.03$ & \\
\hline
$\kappa_\nu$ & $\mu_{\nu_e}$ & $\rho_{\nu_e}+\rho_{\bar{\nu}_e}$ & $\mu_{\nu_e}$ & $\rho_{\nu_e}+\rho_{\bar{\nu}_e}$ & $\mu_{\nu_e}$ & $\rho_{\nu_e}+\rho_{\bar{\nu}_e}$ \\
\hline
$1$ & $0$ & $0.0359$ & $0.318$ & $0.0848$ & $0.903$ & $0.191$ \\
$0.5$ & $0$ & $0.0369$ & $0.299$ & $0.0863$ & $0.829$ & $0.184$ \\
$0$ & $0$ & $0.0379$ & $0.277$ & $0.0879$ & $0.732$ & $0.173$ \\
\hline
\end{tabular}
\label{Table:mu_ell}}
\end{center}
\end{table}

To understand the relation of $g_*^{\ell=0.03, \kappa_\nu=0} < g_*^{\ell=0.03, \kappa_\nu=1}$ more concretely. We give a specific example. Table \ref{Table:mu_ell} shows the numerical solution of the chemical potential $\mu_{\nu_e}$ and energy density $\rho_{\nu_e}+\rho_{\bar{\nu}_e}$ of the electron  neutrinos at $T=0.5$ GeV. The behaviour of the energy density $\rho_{\nu_e}+\rho_{\bar{\nu}_e}$ in the Table \ref{Table:mu_ell} and the behaviour of the relativistic effective degrees of freedom $g_*$ in the Figure \ref{Fig:gast_T} are same. The chemical potential $\mu_{\nu_e}$ is no longer constant with $\kappa_\nu$ even if the lepton number $\ell=0.01$ or $\ell=0.03$ are constant; e.g., for $\ell=0.03$, $(\mu_{\nu_e}^{\kappa_\nu=1}, \mu_{\nu_e}^{\kappa_\nu=0.5}, \mu_{\nu_e}^{\kappa_\nu=0})=(0.903,0.829,0.732)$, so that $\mu_{\nu_e}^{\kappa_\nu=1} > \mu_{\nu_e}^{\kappa_\nu=0.5} > \mu_{\nu_e}^{\kappa_\nu=0}$. 

We can appreciate these behaviour of the chemical potential of neutrinos ($\mu_{\nu_e}^{\kappa_\nu=1} > \mu_{\nu_e}^{\kappa_\nu=0.5} > \mu_{\nu_e}^{\kappa_\nu=0}$) by the following qualitative or semi-quantitative analysis. At GeV scale, all particle species $i$, included neutrinos, are approximately regarded as the ultra-relativistic particles ($m_i \ll T$). For massless particles, the net number density of particle species $i$ is obtained as
\begin{eqnarray}
(n_i-n_{\bar{i}})_{\kappa=1}= \frac{g_i T^3}{6\pi^2} \left[ \pi^2 \left(\frac{\mu_i^{\kappa=1}}{T} \right) + \left( \frac{\mu_i^{\kappa=1}}{T} \right)^3 \right],
\end{eqnarray}
for pure Fermi-Dirac distributions or
\begin{eqnarray}
(n_i-n_{\bar{i}})_{\kappa=0}=\frac{g_iT^3}{\pi^2}\left(e^{\mu_i^{\kappa=0}/T}-e^{-\mu_i^{\kappa=0}/T}\right),
\end{eqnarray}
for pure Maxwell-Boltzmann distributions \cite{Kolb1990,Kolb1980NPB}, where $\mu_i^{\kappa=1}$ and $\mu_i^{\kappa=0}$ denote the chemical potentials in pure Fermi-Dirac and Maxwell-Boltzmann distribution functions, respectively. For sake of simplicity, we assume that the neutrino $\nu$ is only particle species in the universe and ignore the effect of other particle species. In this case, the lepton flavour conservation law to be $s\ell_\nu=(n_\nu-n_{\bar{\nu}})_{\kappa=1}$ for pure Fermi-Dirac neutrino or $s\ell_\nu=(n_\nu-n_{\bar{\nu}})_{\kappa=0}$ for pure Maxwell-Boltzmann neutrino. The equation of $(n_\nu-n_{\bar{\nu}})_{\kappa=1}=(n_\nu-n_{\bar{\nu}})_{\kappa=0}$, or equivalently,
\begin{eqnarray}
\pi^2 \left(\frac{\mu_\nu^{\kappa=1}}{T} \right) + \left( \frac{\mu_\nu^{\kappa=1}}{T} \right)^3 =6\left(e^{\mu_\nu^{\kappa=0}/T}-e^{-\mu_\nu^{\kappa=0}/T}\right),
\end{eqnarray}
should be satisfied for constant $s\ell_\nu$. Thus, we obtain $\mu_\nu^{\kappa=1} > \mu_\nu^{\kappa=0}$ if $s\ell_\nu$ is constant. Indeed, $(\mu_{\nu_e}^{\kappa=1},\mu_{\nu_e}^{\kappa=0}) = (0.318, 0.277)$ for $\ell=0.01$ and $(\mu_{\nu_e}^{\kappa=1},\mu_{\nu_e}^{\kappa=0}) = (0.903, 0.732)$ for $\ell=0.03$ with $T=0.5$ in the Table \ref{Table:mu_ell} approximately satisfy the equation of $(n_\nu-n_{\bar{\nu}})_{\kappa=1}=(n_\nu-n_{\bar{\nu}})_{\kappa=0}$.

The energy density for massless particles is given by
\begin{eqnarray}
(\rho_i+\rho_{\bar{i}})_{\kappa=1}=\frac{\pi^2g_i T^4}{15}\left[\frac{7}{8} + \frac{15}{4}\left(\frac{\mu_i^{\kappa=1}}{\pi T}\right)^2 + \frac{15}{8}\left(\frac{\mu_i^{\kappa=1}}{\pi T}\right)^4 \right],
\label{Eq:rho_kappa1}
\end{eqnarray}
for pure Fermi-Dirac distributions or
\begin{eqnarray}
(\rho_i+\rho_{\bar{i}})_{\kappa=0}=\frac{3g_i T^4}{\pi^2}\left(e^{\mu_i^{\kappa=0}/T}+e^{-\mu_i^{\kappa=0}/T}\right),
\label{Eq:rho_kappa0}
\end{eqnarray}
for pure Maxwell-Boltzmann distribution \cite{Kolb1990,Kolb1980NPB}. If the chemical potential $\mu_\nu$ is constant with $\kappa_\nu$, $\mu_\nu^{\kappa=0}=\mu_\nu^{\kappa=1}$, then the energy density of the neutrino in the case of pure Maxwell-Boltzmann case is lager that it in the case of pure Fermi-Dirac case $(\rho_\nu+\rho_{\bar{\nu}})_{\kappa=0} > (\rho_\nu+\rho_{\bar{\nu}})_{\kappa=1}$ as our naive expectation. However, in the case of $\mu_\nu^{\kappa=0} < \mu_\nu^{\kappa=1}$, the different relation of $(\rho_\nu+\rho_{\bar{\nu}})_{\kappa=0} \le (\rho_\nu+\rho_{\bar{\nu}})_{\kappa=1}$ is also allowed as shown in Table \ref{Table:mu_ell}. 

\subsection{MeV scale}

\begin{figure}[t]
\begin{center}
\includegraphics[width=9.5cm]{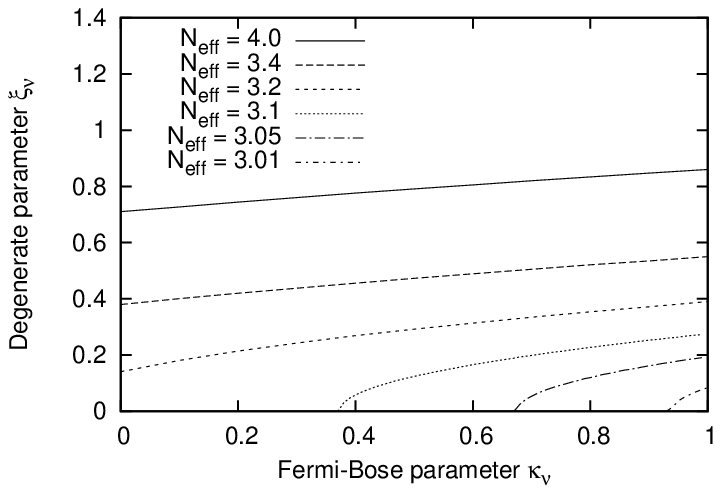}
\caption{The dependence of the degenerate parameter of neutrinos $\xi_\nu$ on the Fermi-Bose parameter $\kappa_\nu$.}
\label{Fig:Neff}
\end{center}
\end{figure}

We also study the bounds on the degenerate parameter of neutrinos $\xi_{\nu_f}=\mu_{\nu_f}/T$ as a function of the Fermi-Bose parameter $\kappa_\nu$ at temperature $T\simeq 1$ MeV (around BBN). These bounds on the degenerate parameters are derived from the effective number of neutrinos $N_{eff}$ at BBN. 

For pure fermionic neutrinos, $\kappa_\nu=+1$, we obtain the effective number of neutrinos as a function of $\xi_{\nu_f}$ as follows:
\begin{eqnarray}
N_{eff}=3+\frac{30}{7}\sum_f \left[ \left( \frac{\xi_{\nu_f}}{\pi}\right)^2+\frac{1}{2}\left( \frac{\xi_{\nu_f}}{\pi}\right)^4\right],
\label{Eq:Neff1}
\end{eqnarray}
in the limit of the instantaneous neutrino decoupling \cite{Stuke2012JCAP} (Including the effect of slight reheating of the neutrinos from early $e^+e^-$ annihilation \cite{Mangano2005NPB}, we obtain $N_{eff}=3.046$ with $\xi_{\nu_f}=0$). On the other hand, the effective number of neutrinos $N_{eff}$ depends on the energy density of the total radiation components $\rho_R$, of the photons $\rho_\gamma$ and of the neutrinos $\rho_\nu=\sum_f \rho_{\nu_f}$ as follows \cite{Mangano2005NPB,Steigman1977PLB}:
\begin{eqnarray}
\rho_R = \left[ 1+\frac{7}{8} \left(\frac{T_\nu}{T}\right)^4 N_{eff} \right] \rho_\gamma = \rho_\gamma + \rho_\nu,
\label{Eq:Neff2}
\end{eqnarray}
where $\rho_\gamma=(\pi^2/30)g_\gamma T^4$ and $g_\gamma=2$. Thus, the effective number of neutrinos $N_{eff}$ is essentially the ratio of the energy density of neutrinos and of photons 
\begin{eqnarray}
N_{eff}=\frac{\rho_\nu}{\frac{7}{8}(\frac{T_\nu}{T})^4\rho_\gamma},
\label{Eq:Neff3}
\end{eqnarray}
and $N_{eff}$ as well as $\rho_\nu$ depends on the degenerate parameters $\xi_{\nu_f}$. We note that if there are some relativistic non-standard model particles, we must include the energy density of these exotic particles on the energy density of neutrinos.

For non-pure fermionic neutrinos, $\kappa_\nu \neq +1$, we still use Eq.(\ref{Eq:Neff3}) as a definition of the effective number of neutrinos.  In the remain of this paper, we perform a rough estimation of the effective number of neutrinos to obtain the rough bounds of the degenerate parameter of neutrinos $\xi_\nu=\xi_{\nu_e}=\xi_{\nu_\mu}=\xi_{\nu_\tau}$ with $\kappa_\nu \neq +1$. The factor $7/8$ in the denominator is due to the Fermi-Dirac statistics of neutrinos, so that it is not certain definition. However, it is sufficient definition for our purpose. In order to recover the value of $N_{eff}=3$ in the case of pure fermionic $\kappa_\nu=+1$ and non-degenerate $\xi_\nu=0$ neutrinos, we take $T=1$ MeV and assume $T_\nu=T$ at this energy scale.

Figure \ref{Fig:Neff} shows the dependence of the degenerate parameter of neutrinos $\xi_\nu$ on the Fermi-Bose parameter $\kappa_\nu$. The six curves in the figure show the numerical solution of $\xi_\nu$ for $N_{eff}=3.01,3.05,3.1,3.2,3.4,4.0$ from bottom to top. Consideration of light element abundance produced at BBN can strongly restrict both chemical potential of neutrinos and the effective number of neutrinos. For example, the recent observational result on the effective number of neutrinos at BBN \cite{Steigman2012AdvHEP} is  $N_{eff}=3.71^{+0.47}_{-0.45}$. On the other hand, in particular, the chemical potentials are bounded \cite{Dolgov2002NPB} as $\vert \xi_\nu \vert <0.07$. Figure \ref{Fig:Neff}  shows that quite large degenerate parameter of neutrinos is required if we try to explain the excess of the effective number of neutrinos, $N_{eff}>3$, without non-standard model particles. In other works, the change of the statistics of neutrinos from Fermi-Dirac to Maxwell-Boltzmann is not sufficient to cover the excess of the effective number of neutrinos. It suggests that, if $N_{eff}>3$ is true, the existence of non-standard model particles as so-called dark radiation are required \cite{Menestrina2012PRD} in the early universe. 

\section{Conclusions}
We have investigated the effect of the presence of non-pure fermionic neutrinos on the relativistic effective degrees of freedom at temperature $0.5 - 500$ GeV. One may expect that the energy density of the universe in the radiation dominant era will increase with deviation of the statistics of neutrinos from pure Fermi-Dirac statistics. In spite of this naive expectation, we have shown that the relativistic degrees of freedom decreases with deviation from pure Fermi-Dirac statistics of neutrinos if there are constant and large lepton asymmetries at GeV scale universe.

The relativistic effective degrees of freedom increases with the energy density and this energy density increases with distribution function of particle species $i$; $f_i=\frac{g_i}{e^{(E-\mu_i)/T}+\kappa_i}$. If the Fermi-Bose parameter $\kappa_\nu$ change continuously from $\kappa_\nu=1$ to $\kappa_\nu \neq 1$, even if the lepton flavour number is constant, the chemical potential $\mu_\nu$ is no longer constant with $\kappa_\nu$. The small $\kappa_\nu$ induces large $f_\nu$ while the small $\mu_\nu$ induces small $f_\nu$. If the effect of $\mu_\nu$ on $f_\nu$ is dominant, the energy density and relativistic effective degrees of freedom are reduced.

We have also studied the bounds on the degenerate parameter of neutrinos $\xi_{\nu_f}=\mu_{\nu_f}/T$ as a function of the Fermi-Bose parameter $\kappa_\nu$ at temperature $T\simeq 1$ MeV (around BBN). It tuned out that the change of the statistics of neutrinos from Fermi-Dirac to Maxwell-Boltzmann is not sufficient to cover the excess of the effective number of neutrinos $N_{eff}>3$. It suggests that the existence of non-standard model particles as so-called dark radiation are required in the early universe. 

In this paper, we have assumed that $0\le \kappa_\nu \le 1$. If we allow the negative case, $\kappa_\nu<0$ (boson like states of neutrinos), the maximum chemical potential should be constrained to $\mu_\nu^{(max)}=m_\nu-T\ln (-\kappa_\nu)$ because $f_\nu$ should be non-negative \cite{Dolgov2005JCAP}. In this case, some new knowledge of chemical potentials of neutrinos as well as the relativistic effective degrees of freedom will be obtained.

\section*{Acknowledgments}
The authors would like to thank the anonymous referees for the comments and suggestions that helped to improve the paper.


\end{document}